# scientific reports

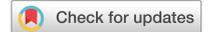

OPEN

# Incremental high average-utility itemset mining: survey and challenges

Jing Chen[1,3,7], Shengyi Yang[2,7], Weiping Ding[4✉], Peng Li[5], Aijun Liu[3✉], Hongjun Zhang[1] & Tian Li[6]

The High Average Utility Itemset Mining (HAUIM) technique, a variation of High Utility Itemset Mining (HUIM), uses the average utility of the itemsets. Historically, most HAUIM algorithms were designed for static databases. However, practical applications like market basket analysis and business decision-making necessitate regular updates of the database with new transactions. As a result, researchers have developed incremental HAUIM (iHAUIM) algorithms to identify HAUIs in a dynamically updated database. Contrary to conventional methods that begin from scratch, the iHAUIM algorithm facilitates incremental changes and outputs, thereby reducing the cost of discovery. This paper provides a comprehensive review of the state-of-the-art iHAUIM algorithms, analyzing their unique characteristics and advantages. First, we explain the concept of iHAUIM, providing formulas and real-world examples for a more in-depth understanding. Subsequently, we categorize and discuss the key technologies used by varying types of iHAUIM algorithms, encompassing Apriori-based, Tree-based, and Utility-list-based techniques. Moreover, we conduct a critical analysis of each mining method's advantages and disadvantages. In conclusion, we explore potential future directions, research opportunities, and various extensions of the iHAUIM algorithm.

Data Mining (DM) refers to a technique for discovering interesting and meaningful data patterns in large databases. This discipline effectively integrates machine learning, statistics, and database systems[1,2] to analyze datasets and discover hidden relationships. ARM[3–6] is a data mining method that is well-known for discovering significant relationships between database items[7–9]. Frequent Pattern Mining (FPM)[10–13], an approach for detecting recurrent patterns in binary datasets[14,15], is widely used in ARM[16–18]. The approach can effectively find the relationships between patterns[16–18] and has been implemented in various real-world problems[19–22].

Two commonly used algorithms for mining patterns in binary databases are Apriori[23] and FP-Growth[24]. Apriori uses a breadth-first search (BFS) algorithm and requires multiple scans of the database. FP-Growth, on the other hand, uses a DepthFirst Search (DFS) algorithm with an FP-tree structure, requiring only two scans of the database. Frequent Itemset Mining (FIM)[25–28] is a well-known research topic aiming to discover frequent itemsets (FI)[24] from a database. However, frequency alone is not always accurate or meaningful in real-world mining scenarios. In a retail market, for example, frequent items may indicate low-profit products, as lower-priced products tend to sell better. Conversely, infrequent itemsets have the potential to generate high profits.

Addressing the limitations inherent in conventional techniques, comprehensive research and successive studies[29–31] have led to proposition and development[32–35] of the HUIM algorithm. Unlike conventional methods that concentrate only on the frequency of itemsets, each item's internal and external utility were considered in HUIM. Utilizing the High-Utility Itemset Mining approach, the utility value of an itemset elevates in correlation with its size, underlined by the quantity of elements within the itemset. Consequently, itemsets of greater length typically yield higher profits, representing a crucial metric for determining high-utility itemsets.

[1]School of Internet of Things, Nanjing University of Posts and Telecommunications, Nanjing 210023, Jiangsu, China. [2]School of Physics and Mechatronic Engineering, Guizhou Minzu University, Guiyang 550025, Guizhou, China. [3]Baotou Teachers' College of Inner Mongolia University of Science and Technology, Baotou 014030, Inner Mongolia, China. [4]School of Information Science and Technology, Nantong University, Nantong 226019, Jiangsu, China. [5]School of Computer Science, Nanjing University of Posts and Telecommunications, Nanjing 210023, Jiangsu, China. [6]School of Computer and Software, Nanjing Vocational University of Industry Technology, Nanjing 210003, China. [7]These authors contributed equally: Jing Chen and Shengyi Yang. ✉email: dwp9988@163.com; laj66261@163.com

Hong et al.[36] proposed the HAUIM approach, which measures the average utility of itemsets based on their length to provide a fair evaluation of itemsets. If the average utility of an itemset meets or exceeds a predetermined minimum threshold, it is identified as a High Average Utility Itemset (HAUI). Consequently, as compared to traditional HUIM, presents a distinct set of challenges that necessitate the development of novel techniques, including downward-closure properties, upper-bound models, pruning strategies, and mining procedures. In their work, Lin et al. introduced a tree-based algorithm called HAUP tree[37] for mining HAUI sets. They leveraged this efficient tree structure to enhance mining performance. Furthermore, to enhance the speed of the mining process, they introduced a projection-based algorithm called PAI[38]. In a subsequent study, Lin et al.[39] devised an innovative data structure known as the Average Utility (AU) list, which efficiently mines HAUI from static databases. This AU list-based approach represents the current state-of-the-art algorithm for mining HAUI.

The efficiency of existing techniques used to detect HAUI in a static database can be compromised when the size of the database undergoes changes. Specifically, when new transactions are introduced, it necessitates reprocessing the entire database in order to update the results. To address this issue, Cheung et al.[40] introduced the concept of FUP (Fast Update) to preserve the discovered frequent itemsets through incremental updates. As the database changes, their framework considers four scenarios in which the updates are handled differently depending on the prescribed methods. The FUP concept has already been applied to ARM[40,41], HUIM[42,43], and HAUIM[44,45]. However, these methods still have the disadvantage of rescanning certain itemsets and requiring additional database scans to obtain these itemsets. To address this challenge, Wu et al.[46] introduced a hierarchical approach that incorporates the pre-large concept in HAUIM for incremental mining. Nevertheless, an important limitation of their model lies in the absence of theoretical evidence supporting the ability of the pre-large concept to effectively preserve the correctness and completeness of the maintained HAUI.

In the past 10 years, researchers have developed over ten algorithms specifically designed for handling dynamic databases in the context of transaction insertion for iHAUIM. The objective of iHAUIM is to identify patterns that meet the minimum utility constraints while continuously inserting new records into the original database. This problem can be considered as a constraint-based mining problem. The development of efficient iHAUIM algorithms is a new research problem because it makes iHAUIM tasks more scalable with respect to database updates.

Real-time processing of data streams has become essential due to the increasing number of applications, including auditors, online clickstreams, and power throughput, that generate data streams that require immediate processing. These data streams are generated rapidly and accumulate in real time, demanding efficient processing methods. To address this, a single scan of the data stream is typically employed to build a data structure, which is then maintained throughout the execution. This approach ensures that newly generated data influences the resulting patterns. When new data is inserted, the data structure is updated and reconstructed to enable efficient mining. Traditional methods used for processing static data, which involve multiple scans of the database and deletion of unwanted items, are not suitable for handling data streams. Instead, techniques like sliding windows[47–49], damping windows[50–52], and landmark windows[53] are employed to effectively handle stream data.

Moreover, in[54], a sliding window model is utilized, alongside a decay factor. MPM[55] and DMAUP[52] are both mining methods aiming to identify high average utility patterns, by employing a damping window concept. Essentially, they analyze recent transactions more heavily than prior ones, but they struggle to function effectively with large databases, especially patterns that frequently occur in recently made transactions. This is due to their tendency to process the entire database each time they encounter a new data stream. In addition to this, each computation of the decay factor is considerably computation-heavy. MPM, being a tree-based method, is unable to store the actual utility of the respective items. This results in consuming a lot of runtime and memory for generating candidate patterns. Plus, verifying candidate patterns for accuracy demands extra database scans, and therefore, it is unsuitable for data stream analysis.

Although the iHAUIM algorithms have been developed, there has been no comprehensive exploration or empirical study to compare their performance. The primary objective of this paper is to provide a comprehensive and in-depth analysis of the notable progress in iHAUIM. The methodologies discussed in this study can serve as valuable insights not only for iHAUIM but also for other data mining tasks, including incremental data mining[56,57] and dynamic data mining[58]. In[57], a dynamic and incremental profit environment is explored, and a unique approach named IncDEFIM is introduced. This method employs strategies like merging transactions, projecting databases, and setting strict upper bounds to minimize the expenses associated with database scans while efficiently removing unproductive item sets. By examining these advancements, this research aims to contribute to the broader field of data mining and inspire further developments in related domains.

This article made three distinct contributions. Firstly, it provided a comprehensive overview of the essential technologies employed in iHAUIM algorithms. Secondly, it conducted a comprehensive review of the latest advancements in this field. Lastly, it identified and emphasized potential areas for future research in data mining. Moreover, this research paper presents a new classification system that integrates contemporary methods for extracting HAUIs from dynamic datasets. As a result, it offers a valuable framework for advanced iHAUIM algorithms, eliminating redundancies in the existing literature. The main contributions of this work are as follows:

1. The paper proposes a classification approach for the most advanced iHAUIM algorithms that includes the most up-to-date information on methodologies for extracting HAUIs from dynamic datasets.
2. Based on the dynamic datasets, we categorize HAUIM algorithms into three types: Apriori-based, Tree-based, and Utility-list-based.

3. The article provides a thorough comparison of the benefits and drawbacks of the most sophisticated iHAUIM algorithms, including metrics such as running time, memory usage, scalability, data structures, and pruning techniques.
4. Furthermore, this paper offers a comprehensive summary and discussion of current iHAUIM techniques. Lastly, it outlines potential research possibilities and key areas for future iHAUIM research.

The structure of this article is as follows: "Preliminaries and problem statement of iHAUIM" section provides an overview of the fundamental concepts and definitions related to iHAUIM. "State-of-the-art algorithms for iHUIM" section classifies and explains iHAUIM approaches based on dynamic datasets, evaluating their advantages and disadvantages. "Summary and discussion" section presents a thorough overview and evaluation of the latest iHAUIM techniques, which highlights potential research directions and opportunities for future advancements in iHAUIM. Lastly, "Conclusion" section concludes the survey, summarizing the key findings and contributions of the article.

## Preliminaries and problem statement of iHAUIM

In this section, we lay the foundation by providing essential preparations and presenting a formal definition of the iHAUIM problem. We will also introduce the symbols that will be used throughout the rest of this paper, as shown in Table 1, and these symbols will be explained in subsequent sections. Below are examples of the original database and item utility table, presented as Table 2 and Table 3, respectively. The original database comprises five transactions, each identified by a transaction identifier (TID) and containing non-redundant items. The internal utility of each item is specified after a colon. Table 3 displays seven items that are present in the original database, represented as I = {a, b, c, d, e, f, g}. The external utility of each item is shown in Table 3.

| Notation | Meaning |
|---|---|
| I | A set of m items,I = {$i_1$, $i_2$, ..., $i_m$}, where each item $i_j$ has a profit value $p_j$ |
| DB | An original quantitative database, DB = {$T_1$, $T_2$, ..., $T_n$}, in which each transaction is a subset of I, with purchase quantities for each item |
| DBn+ | A set of new transactions, DBn = {$t_1$, $t_2$, ..., $t_q$}, in which each transaction includes a subset of items, with purchase quantities |
| TID | Each transaction $T_n \in D$ has a unique transaction identifier (TID) |
| X | A k-itemset containing k distinct items {$i_1$, $i_2$, ..., $i_k$} |
| u($i_j$, $T_p$) | The utility of an item $i_j$ in a transaction $T_p$ |
| u($T_p$) | The sum of the utilities of items in a transaction $T_p$ |
| $tu_{DB}$ | The total utility $tu_{DB}$ of a database DB |
| au(X,$T_p$) | The average utility of X in $T_p$ |
| au(X) | The average utility of X in DB |
| HAUI | High-average-utility Itemset |
| mu($T_p$) | The maximum utility of transaction $T_p$ |
| auub($i_j$) | The average-utility upper-bound (AUUB) of item $i_j$ |
| $HAUUBI_{DB}$ | High average-utility upper bound itemset |
| $PAUUBI_{DB}$ | Pre-large average-utility upper-bound itemset |
| $HAUI_{UDB}$ | An itemset X is classified as an $HAUI_{UDB}$ in the updated (DB + DBn) database |

**Table 1.** Notation.

| Transaction | U |
|---|---|
| a:4, b:1, c:2, d:1 | 33 |
| b:4, d:3, e:1, g:11 | 69 |
| a:1, c:4, f:3 | 40 |
| c:1, d:2, e:2 | 41 |
| a:2, d:1, f:5 | 49 |

**Table 2.** An example database (DB).

| Item | a | b | c | d | e | f | g |
|---|---|---|---|---|---|---|---|
| Profit | 3 | 5 | 4 | 8 | 11 | 7 | 2 |

**Table 3.** Unit profits of items.

The minimum high average utility upper-bound threshold δ and the lower-bound threshold $δ_L$ are set based on the user's preference (positive integers). Below are commonly used definitions for incremental high average utility pattern mining[44,59,60], sliding window[47,49,61], and dampened window models[52,55], derived from the provided original database and item utility table.

**Definition 1** Item utility[62]. The utility of an item $i_j$ in a transaction $T_p$ is represented as $u(i_j, T_p)$ and is computed as the product of its internal utility in transaction $T_p$, denoted as $iu(i_j, T_p)$[62], and its external utility $eu(i_j)$.

$$u(i_j, T_p) = iu(i_j, T_p) \times eu(i_j) \tag{1}$$

For instance, in Table 2, the item utility of 'a' in $T_1$ is calculated as $u(a, T_1) = 3 \times 4 = 12$.

**Definition 2** Transaction utility[63]. The transaction utility of[63] $T_p$ is indicated and computed as follows[52].

$$u(T_p) = \sum_{i_j \in T_p} u(i_j, T_p) \tag{2}$$

For instance, in Table 2, the transaction utility of $T_1$ is calculated as $u(T_1) = u(a, T_1) + u(b, T_1) + u(c, T_1) + u(d, T_1) = 12 + 5 + 8 + 8 = 33$.

**Definition 3** Total utility[64]. The total utility ($tu_{DB}$) of a database DB is defined as follows:

$$tu_{DB} = \sum_{T_p \in DB} u(T_p) \tag{3}$$

As an example, the total utility in the illustrated case of Table 2 is computed as $tu_{DB} = 33 + 69 + 40 + 41 + 49$ (= 232).

**Definition 4** Average utility[62]. The average utility of item X in transaction $T_p$, denoted as $au(X, T_p)$, is calculated by dividing the sum of item utilities in X by the length of X[61], |X|.

$$au(X, T_p) = \frac{\sum_{X \subseteq T_p \wedge i_j \in X} u(i_j, T_p)}{|X|} \tag{4}$$

**Definition 5** Itemset Average utility. The average utility of X in the database ($au(X)$) is determined by summing up the average utilities of X in all transactions present in the database DB[61].

$$au(X) = \sum_{T_p \in DB \wedge X \subseteq T_p} au(X, T_p) \tag{5}$$

For instance, in Table 2, the average utility of 'ac' in the database is calculated as $au(ac) = au(ac, T_1) + au(ac, T_3) = 10 + 9.5 = 19.5$.

**Definition 6** HAUI. An itemset is categorized as a HAUI if its au satisfies[65]:

$$HAUI \leftarrow \{X au(X) \geq TU_{DB} \times \delta\} \tag{6}$$

For example, if δ is 8%, then itemset a, c is a HAUI since $au(a, c) = 19.5 \geq 2320.08 = 18.56$.

**Definition 7** Maximum utility[66]. The maximum utility of[66] transaction $T_p$ is notated as follows:

$$mu(T_p) = u(i_j, T_p) \tag{7}$$

For instance, in Table 2, the maximum utility of $T_1$ is calculated as $mu(T_1) = 12$.

**Definition 8** AUUB[55]. For an item $i_j$, the AUUB of $i_j$ is as follows:

$$auub(i_j) = \sum mu(T_p), \text{ where } i_j \in T_p \text{ and } T_p \in DB \tag{8}$$

For instance, in Table 2, the AUUB of a is $auub(a) = mu(T_1) + mu(T_3) + mu(T_5) = 12 + 21 + 35 = 68$.

**Property 9** *DC, Downward closure property of AUUB[46].*

*According to the downward closure property of AUUB[46], if an itemset Y is a superset of itemset X[46], denoted as $Y \supseteq X$, the following formula (9) can be obtained.*

$$auub(X)_{DB} \geq auub(Y)_{DB} \tag{9}$$

*Hence, if $auub(X)_{DB} \geq tu_{DB} \times \delta$ then $auub(Y)_{DB} \leq auub(X)_{DB} \leq tu_{DB} \times \delta$ is satified for any superset of $X$[46].*

**Definition 10** HAUUBI[46]. For the dataset DB, if an itemset X is a HAUUBI$_{DB}$, it should satisfy the following condition as[46]:

$$\text{HAUUBI}_{DB} \leftarrow \{\text{auub(X)}_{DB} \geq \text{tu}_{DB} \times \delta\} \tag{10}$$

**Definition 11** PAUUBI. Itemset X is a PAUUBIDB in the initial database[62]:

$$\text{PAUUBI}_{DB} \leftarrow \{X | \text{tu}_{DB} \times \delta_L \leq \text{auub(X)}_{DB} \leq \text{tu}_{DB} \times \delta\} \tag{11}$$

For instance[62], suppose $\delta$ the and $\delta_L$ are respectively set as 13% and 8%. The itemset (ce) is PAUUBI with an auub of 26, which lies between $\delta_L$ (= 232 × 8%)(= 18.56) and $\delta$(= 232 × 13%) (= 30.16).

**Definition 12** The condition of HAUI$_{UDB}$.In the updated (DB + DBn) database, in Table 4, an itemset X qualifies as a HAUI$_{UDB}$ if it meets the following conditions[46]:

$$\text{HAUI}_{UDB} \leftarrow \{X | \text{au(X)}_{UDB} \geq (\text{TU}_{DB} + \text{TU}_{DBn+}) \times \delta\} \tag{12}$$

where au(X)$_{UDB}$ indicates the new average-utility of X,TU$_{DB}$ and TU$_{DBn+}$ are respectively the transaction utility in DB and DBn +, and $\delta$ is the upper bound of utility threshold[46].

## State-of-the-art algorithms for iHUIM

In recent years, a considerable number of iHAUIM (Insert-based High Average Utility Itemset Mining) techniques have been developed to handle dynamic databases involving transaction insertions. So far,a total of 19 iHAUIM algorithms have been proposed, as shown in Fig. 1, which can be classified into three main categories: apriori-based, tree-based, and utility-list-based methodologies. In the upcoming sections, we will evaluate the strengths and weaknesses of each algorithm, as indicated in Table 5, with the primary aim of mining itemsets that exhibit high average utility during transaction updates.

The traditional HAUIM algorithm is only applicable to static datasets. However, if the dataset undergoes record updates, the static techniques necessitate processing all the data from the start to extract HAUI. Consequently, this results in high time and memory consumption.

| (A) DB1+ | (A) DB2+ | | |
|----------|----------|-----|----------|
| TID | Contents | TID | Contents |
| T6 | a:3, b:5, c:3, d:1 | T8 | a:3,b:5,d:1,f:4 |
| T7 | a:4, c:2, e:1, f :1, g:5 | T9 | a:5,b:6,c:1,e:1,f:3 |

**Table 4.** Additional DB1+, DB2+ and the updated MU table.

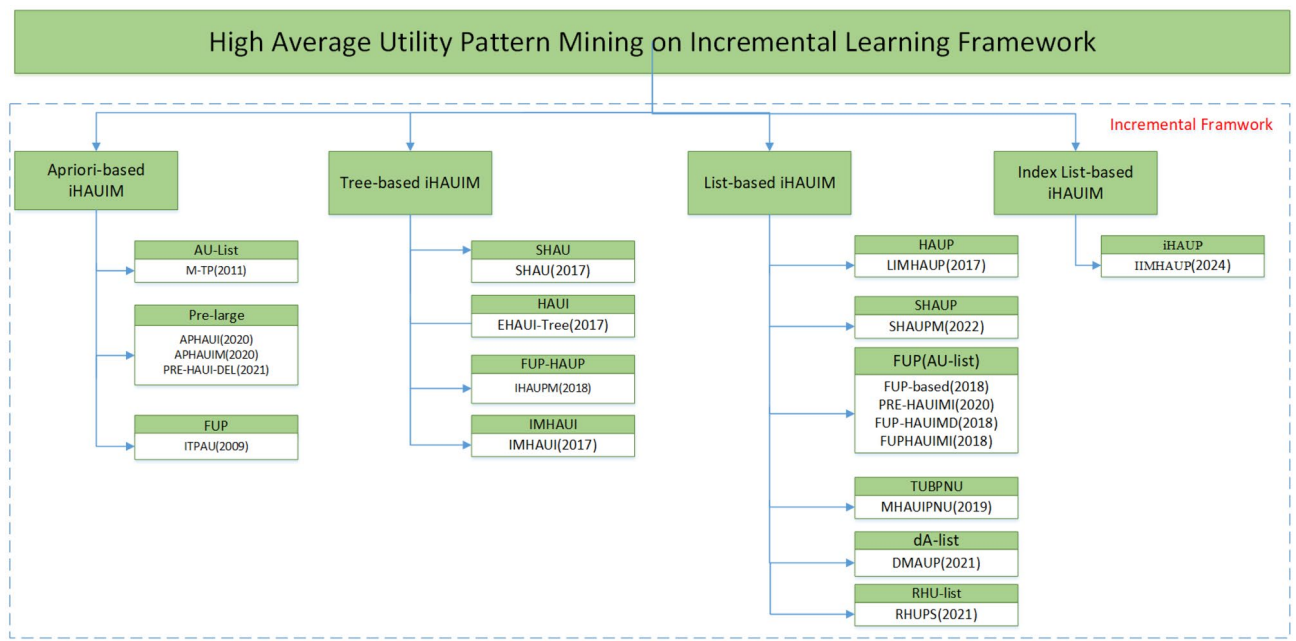

**Figure 1.** Classification of iHAUIM Algorithms.

| Algorithm | IM | DCP | TSA | HDS | PNU | PpP | TD | IU |
|-----------|-----|-----|-----|-----|-----|-----|-----|-----|
| ITPAU | √ | √ | √ | √ | | | | √ |
| M-TP | √ | √ | | √ | | √ | | √ |
| SHAU | | √ | | √ | | | | √ |
| EHAUI | √ | √ | | | | √ | | √ |
| IMHAUI | | | | | | | | √ |
| FUP-based | √ | | | √ | | | | √ |
| MAM | √ | √ | | | | | | √ |
| IHAUPM | | | | | √ | | | √ |
| FUPHAUIMI | | | | | | | | √ |
| FUP-HAUIMD MHAUIPNU | √ | | | | | √ | √ | √ |
| PRE-HAUIMI | √ | | | | | √ | | √ |
| LIMHAUP | | | | | | √ | | √ |
| APHAUI | | | | √ | | | √ | |

**Table 5.** Algorithm advantage. *IM* incremental maintenance, *DCP* downward closure property, *TSA* two-stage algorithm, *HDS* handling data streams, *PNU* positive and negative utilities, *PpP* pre-processing and pruning, *TD* transaction deletion, *IU* incremental updates.

## Apriori-based iHAUIM

Based on the Fast Update (FUP) concept[40], the TPAU[67] algorithm discovered HAUI from dynamic datasets that change with the insertion of new records. FUP records the previously frequent large itemsets and their counts for use in the maintenance process. when a new transactions are added, the FUP(Frequent Utility Pattern) algorithm generates candidate 1-itemsets. Subsequently, the candidate itemsets are compared with the previous itemsets in order to classify them into the following four cases:

Case 1:   The itemset is large in both the original database and the newly added transactions, resulting in its categorization as large in both domains.
Case 2:   The itemset is large in the original database but not in the newly inserted transactions.
Case 3:   Although not considered large in the original database, the itemset demonstrates significance in the newly inserted transactions.
Case 4:   The itemset does not meet the threshold for being deemed large in either the original database or the newly inserted transactions.

The suggested algorithm adopts an approach similar to Apriori to systematically explore the layers of HAUI. To optimize the search process, it employs early pruning techniques to discard low-utility itemsets. The algorithm leverages the downward closure property in a two-stage process, enabling it to generate a reduced set of candidate items at each level. In the first stage, an overestimated itemset is obtained using an average utility upper bound. In the second stage, an actual average utility value is computed, considering a high upper bound. Through these steps, the algorithm efficiently extracts HAUIs from incremental transaction datasets, enhancing its mining capabilities. Afterwards, the modified itemsets are categorized into four groups based on their characteristics, and whether their count difference in the modified records is positive, negative, or zero. Each group is then subjected to its specific processing approach.

The M-TP[59] proposes a two-stage record modification maintenance method, aimed at mining HAUI from updated datasets. To begin, this approach calculates the count difference by comparing the AUUB (Average Utility Upper Bound) of each modified itemset before and after modification. Then, the modified itemsets are divided into four parts based on their characteristics. This classification is determined by whether they are HAUUBI (High Average Utility Upper Bound Itemsets) in the original dataset and whether their count difference in the modified records is positive or negative (including zero). Each part is then subjected to its specific processing approach. The M-TP algorithm reduces the time required to reprocess the entire updated dataset. In the original dataset, the itemsets are larger in the first two cases, and smaller in the last two cases. Conversely, the first and third cases exhibit a positive count difference, while in the modified records, the count difference turns negative or remains zero in the second and fourth cases. Lan et al.[59] proposed four cases of modifying records from existing datasets in Fig. 2.

In contrast to conventional approaches, the algorithm[59] reduced the time required for the entire dataset updating time. In terms of runtime, the M-TP algorithm demonstrates superior performance to the Batch TP algorithm across different minimum average utility thresholds[68].

The algorithm[69] is proposed to handle transaction deletions in dynamic databases using the pre-large concept on HAUIM, called PRE-HAUI-DEL. The pre-large concept is used as a buffer to reduce the number of database scans, particularly during transaction deletions, and its overview is illustrated in Fig. 3. Additionally, two upper bounds are established in the algorithm to early prune unpromising candidates, which can accelerate computation costs. Compared to Apriori-like models, PRE-HAUI-DEL excels in efficiently mining high-average utility items in updated databases. In addition, the developed algorithm also uses the LPUB upper bound model, which can significantly reduce the number of candidate items that need to be checked in the search space. Compared to the general model that updates discovered knowledge Using batch processing mode, our designed PRE-HAUI-DEL can effectively maintain the discovered HAUI without the need for multiple database scans, as illustrated in Figs. 4 and 5. This not only reduces computational costs but also correctly and completely maintains knowledge about HAUI.

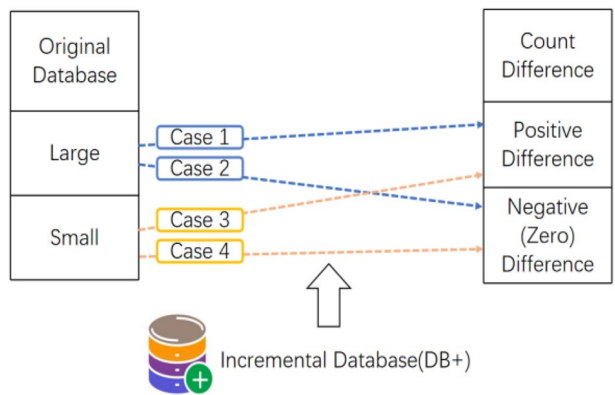

**Figure 2.** Four cases when records are modified from an existing dataset.

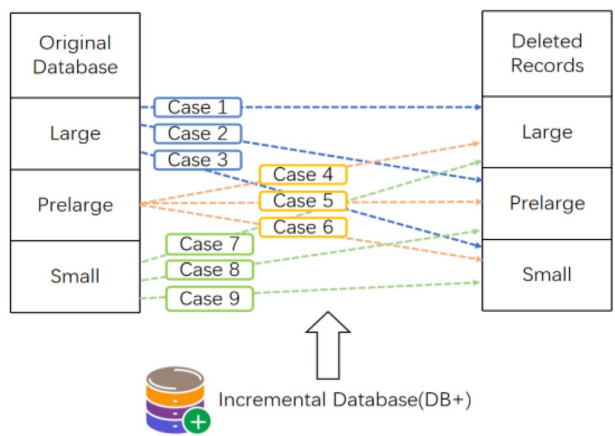

**Figure 3.** Nine cases of the pre-large concept.

In[65], this article introduces the APHAUI algorithm, a HAUP (High-Utility Association Pattern) algorithm based on Apriori, capable of effectively mining HAUI from dynamic datasets. This algorithm follows an Apriori-like approach[23] and employs the pre-large concept[56] to reduce the search space and proactively prune less promising candidate items, revealing promising itemsets during maintenance. The final results of cases 1, 5, 6, 8, and 9 remain unaffected. Moreover, the amount of information discovered in cases 2 and 3 can be reduced, while some new information might emerge in cases 4 and 7. As shown in Fig. 6,the pre-large concept can easily handle itemsets in cases 2, 3, and 4. The authors devised two upper bounds, namely Partial Upper Bound (pub) and Lower pub (lpub), to enhance the efficiency of the mining process. The pub serves as astringent upper limit that reduces the size and upper utility bound of promising itemsets. A High pub itemset (pubi) with greater utility than pub was developed.

Furthermore, the algorithm introduces a subset named lpubi (Lead-pubi) as a part of pubi, capable of further reducing the candidate itemset for subsequent mining processes. Despite the algorithm generating both pubi and lpubi itemsets, the applicability of lpubi is constrained compared to pubi. Lead-pubi contributes to reducing the count of candidate items. Additionally, a formula is employed to curtail unnecessary dataset scans. Lastly, the introduction of a linked list ensures that each transaction is scanned at most once, thereby minimizing the frequency of dataset scans during the update process.

The algorithm begins by scanning the input dataset, followed by the dynamic processing flow of the APHAUI method. By employing a designed re-scanning threshold, it can automatically determine the update pace of the incremental dataset, enhancing mining efficiency. During the algorithm's execution, two upper bounds, pub, and lead-pub, along with two itemsets, pubi and lead-pubi, are used to reveal the complete set of HAUIs within the transaction dataset. This algorithm not only demonstrates strong performance but also holds significant potential in real-time scenarios.

---

**Algorithm 1** PRE-HAUI-DEL[77]

> **Input:** $D$,original transaction dataset.
> $S_u$,upper-bound utility threshold.
> $S_l$,lower-bound utility threshold.
> **Output:** $P$,set of pre-large itemsets.
> $H$,set of HAUI.
> initialize H = $\phi$,P = $\phi$;
> calculate the threshold $\gamma$ for D to perform re-scan process;
> **for** each transaction t in D **do**
>   **for** each item i in t **do**
>     calculate auub(i) and au(i);
>     **if** $au(i) \geq S_u \times TU^D$ **then**
>       put i in H;
>     **else if** $au(i) \geq S_l \times TU^D$ **then**
>       put i in P;
>     **end if**
>     **if** $auub(i) \geq S_u \times TU^D$ **then**
>       put i in pubis and lead-pubis;
>     **end if**
>   **end for**
> **end for**
> **while** pubis $\neq \phi$ and lead-pubis $\neq \phi$ **do**
>   initialize C = $\phi$;
>   **for** each transaction t in D **do**
>     generate candidate itemsets by combining e and the
>     itemsets in pubis $\longrightarrow$ C
>   **end for**
>   initialize pubis $\longleftarrow\phi$,lead-pubis $\longleftarrow\phi$
>   **for** each transaction t in D **do**
>     calculate all of necessary utility information for each
>     itemset in t;
>   **end for**
>   **for** each candidate itemset c in C **do**
>     **if** $pub_c \geq S_l \times TU^D$ **then**
>       put c in pubis;
>     **end if**
>     **if** $lpub_c \geq S_l \times TU^D$ **then**
>       put c in lead-pubis;
>     **end if**
>     **if** $au(c) \geq S_u \times TU^D$ **then**
>       put c in H;
>     **else if** $au(c) \geq S_l \times TU^D$ **then**
>       put c in P;
>     **end if**
>   **end for**
> **end while**
> return P,H,$\gamma$;

**Figure 4.** PRE-HAUI-DEL.

Previous HAUM algorithms processed dynamic datasets using batch processing. As a result, the APHAUIM[46] incurred costs in terms of past computations and the discovery of pattern information. To address this issue, the concept of FUP (Frequent Update Pattern) was introduced[40] for real-time pattern discovery and storage of pattern information. However, this requires rescanning the dataset to acquire the latest information. In[70], a new model called Apriori-based Potential High Utility Itemset Mining (APHAUIM) is proposed, which effectively reveals potential high utility patterns from uncertain databases in industrial IoT by maintaining two item sets (phps and plhps) using two tight upper-bound values (pub and lead-pub), while ensuring the completeness and correctness of the mining results.

Based on the concepts of pre-large[56,58] and the Apriori method[23], a new algorithm called APHAUIM is introduced to mine HAUI from incremental transaction datasets. PAUBI is introduced to retain promising HAUBI. PAUBI acts as a buffer to minimize the rescans needed for checking whether a small itemset evolves into a large itemset. An overview of the pre-large concept is depicted in Fig. 6.

Compared with the benchmark FUP-based HAUIM algorithm[67], the designed algorithm is better suited for streaming environments in dynamic datasets. However, a limitation lies in the fact that, similar to the benchmark method, the proposed algorithm also incurs a considerable amount of rescanning time. This is because locating itemsets in the buffer to update the insertion process requires additional time. Therefore, selecting appropriate thresholds is a topic of significant importance.

## Tree-based iHAUIM

The SHAU[61] introduces an effective algorithm named SHAU for analyzing time-sensitive data in terms of significance. The algorithm employs the HAUPM algorithm based on sliding windows to process data streams. The

**Algorithm 2** Maintenance process of PRE-HAUI-DEL[77]

**Input:** $d$,deleted transaction dataset.
$P$,pre-large itemsets.
$H$,high average-utility itemsets.
$\gamma$,re-scan threshold.
$S_u$,upper-bound utility threshold.
$S_l$,lower-bound utility threshold.
**Output:** $P$,updated pre-large itemsets.
$H$,updated HAUI.
$\gamma$,updated re-scan threshold.
set s is the size of d;
$\gamma = \gamma$ - s;
**if** $\gamma \geq 0$ **then**
    perform the complete re-scanning process(Algorithm 2)
    to obtain $P^{'},H^{'}$;
    $S \leftarrow \sim (P^{'} \bigcup H^{'}) \cap (P \cup H))$
    calculate necessary utility information for S in d; update
    the utility value of $P$, $H$ from $P,H,P^{'},H^{'},S$
**else**
    generate D $\leftarrow$ D - d;
    re-scan the updated dataset D by performing Algorithm
    2;
    update the utility value of $P,H,\gamma$;
**end if**
**return P,H,$\gamma$;**

**Figure 5.** Maintenance process of PRE-HAUI-DEL.

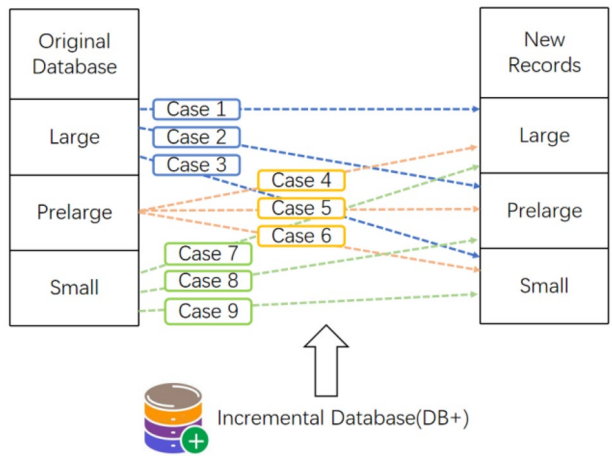

**Figure 6.** Nine cases of the pre-large concept.

HAUPM algorithm considers only new data during the pattern mining process for discovering data streams. As the algorithm is based on the concept of sliding windows[71-74], it divides the data stream into multiple blocks or batches. The concept of sliding windows for data streams was initially proposed by Yun et al.[61].

The SHAU algorithm employs a novel SHAU tree structure. In this tree, each node consists of three elements: the first element stores the tid that includes the item, the second element is used to store the recent auub data information of the data stream batch by batch, and the third element is a link pointing to another node with the same tid. The auub of different items in the data stream is stored in the header table of the SHAU tree. Additionally, the efficiency of SHAU is enhanced by utilizing a new strategy called RUG.

The EHAUI-tree algorithm[75] is proposed as an improved iteration of the HAUI-tree algorithm[76]. The primary objective is to enhance mining efficiency and reduce memory consumption. The algorithm aims to mine by adding new transactions instead of restarting the dataset. It utilizes the downward closure property and employs an index table structure. This innovative approach enhances computational efficiency while simultaneously reducing memory requirements. In addition, the algorithm introuces a bit-array structure to compute utility values more efficiently. However, the algorithm performs poorly on large datasets or small thresholds.

In[45], a new approach called IHAUPM is proposed for handling frequent transaction insertions in updated datasets. The algorithm leverages an adapted FUP concept to efficiently integrate prior information and update the results when new information is discovered during updates. The newly inserted transactions are categorized into four distinct cases, considering the occurrence frequency of the original dataset and the newly inserted transactions. This categorization ensures effective handling of different scenarios and minimizes repetition during the updating process. In cases where the itemset is the original dataset or the HAUUBI in the new insert, it remains a HAUUBI, while in cases where it is not, it remains non-HAUUBI.

For cases where it is necessary to determine whether the itemset is actually a HAUUBI based on existing information or to rescan the original dataset, the algorithm employs a compressed HAUP tree data structure to store and utilize the required information. This approach requires minimal scanning of the original dataset and is highly efficient while preserving the count of prefix items processed in each node of the tree.

This article[60] proposes an algorithm called IMHAUI, which is based on the IHAUI-tree and uses node sharing to preserve the information of the incremental dataset, thereby addressing the problem of adding new data to the dataset which may cause the number of items to exceed or fall below the minimum support threshold. Each time new data is added, node sharing undergoes reconstruction. To achieve this, transactions within the dataset are sorted in descending order based on their AUUB values. During the reconstruction process, each path is rearranged in decreasing order of the optimal AUUB value. To maintain compactness, a path adjustment technique is utilized[77]. Additionally, the algorithm preserves the AUUB value of each itemset by maintaining a header table. Subsequently, the mined tree is examined to access candidate itemsets, and their actual average utility is computed during the candidate validation phase.

FIMHAUI based on mIHAUI-tree, to address the problems of time-consuming candidate itemset generation and expanding search space by determining the upper limit value caused by IMHAUI[60]. The algorithm performs a singlescan of the dataset to extract information from HAUI. It stores transaction information in each node of IHAUI-Tree, which completely overlaps with the path from the root to that node, and thus only saves the necessary information in the leaf nodes of mIHAUI-Tree. Initially, all transactions are inserted into an empty mIHAUI-tree in a sequential order based on alphabetical order. Subsequently, the path adjustment method proposed in[60] is to adjust the paths in order to enhance the sharing efficiency of nodes within the mIHAUI-tree. The algorithm uses data set projection and merge techniques to efficiently find itemsets. mIHAUI-tree introduces a novel approach by directly obtaining the projected data set for candidate itemsets, eliminating the need for generating conditional patterns and local trees. Additionally, a transaction merge technique identifies identical transactions in dictionary order within one scan. In contrast to the IHAUI-tree, the proposed algorithm offers not only time savings but also a reduction in repetition. However, the performance of the algorithm is unsatisfactory when applied to large datasets or small thresholds.

In[55] the MAMs algorithm was designed to effectively analyze time-sensitive data which was applicable to data streams and employs an exponential damping window model and pattern growth methods. Furthermore, the algorithm considers the temporal aspect of the provided data to acquire pertinent and current pattern knowledge. The algorithm employs DAT structure and TUL to handle dynamic data streams. As new data is inserted into a transaction, the algorithm constructs a DAT data structure and incorporates average utility information. This procedure persists until a user-initiated mining request is encountered. Upon receiving such a request, the MPM algorithm follows the pattern growth approach on the dataset.

The common goal of these algorithms is to enhance the efficiency of data mining, reduce memory consumption, and adapt to the dynamic nature of data. The SHAU algorithm utilizes the HAUPM algorithm based on sliding windows to process data streams, employing the SHAU tree structure to store itemset information from the data stream and enhancing efficiency through the RUG strategy. The EHAUI-tree algorithm, as an improved version of the HAUI-tree algorithm, and the IHAUPM algorithm, introduce new methods for handling frequent transaction insertions in updated datasets. The FIMHAUI algorithm, based on the mIHAUI-tree, addresses the time-consuming generation of candidate itemsets and the expansion of search space in IMHAUI. These algorithms share a common challenge in that they attempt to optimize the mining process through various data structures and strategies to accommodate the dynamic changes and time sensitivity of data. However, they may encounter performance issues when dealing with large datasets or small thresholds, indicating that further optimization and improvement may be necessary in practical applications.

## List-based iHAUIM

To address the issue of inadequate performance in mining advanced association rules in dynamic environments, Wu et al. proposed an algorithm[44] to update the obtained advanced association rules using transaction insertion. The proposed algorithm builds upon the AU list[39] and incorporates the concept of FUP (Frequency Upper Bound)[40] to enhance its performance. To adapt and update advanced association rules with transaction insertion, the proposed algorithm employs a two-stage approach. In the initial stage, the 1-HAUUBI set is derived from the original dataset. Subsequently, an AU list is constructed from the 1-HAUUBI set, facilitating subsequent processing. In the second stage, the algorithm efficiently handles transaction insertion by dividing the HAUUBI set into four partitions based on the FUP (Frequency Upper Bound) criterion. This partitioning strategy minimizes repetition and enhances efficiency during the updating process. The proposed algorithm, as described in[44], presents four distinct cases for handling transaction insertion, as illustrated in Fig. 7.

In each case, the algorithm preserves the HAUUBI set for each partition, with the exception of non-advanced association rules in case 4. These non-advanced items are excluded from the HAUUBI set during dataset updates, as they do not qualify as advanced association rules. This approach effectively reduces redundancy in the algorithm, as illustrated in Fig. 8. The updateADD and updateDEL methods are used for adding and deleting items in the AU-list structure, respectively. The updateADD function can easily update the auub value of the itemsets based on the AU-list structure. As for the updateDEL function, it can directly remove the unpromising itemsets based on the AU-list structure after the database has been updated.

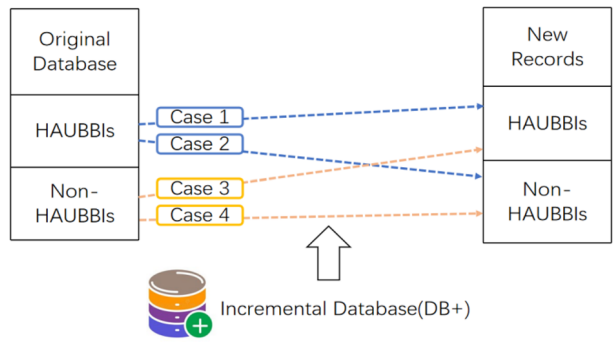

**Figure 7.** Four cases of the proposed algorithm with transaction insertion.

---

**Algorithm 3** Proposed FUP-based[43]

**Input:** $D$,a quantitative database;
  $ptable$,a profit table;
  $TU^D$ the total utility in $D$;
  $\delta$,the minimum high average-utility threshold;
  $AUL$,the built AU-list form $D$;
  $d$,a set of inserted transactions.
**Output:** the sets of HAUIs and HAUUBIs.
set HAUUBIs.U⟵ **null**;
calculate $TU^d$ in d;
**for** each $i_j \in$ d **do**
  calculate auub($i_j$);
  **if** auub($i_j$)$^d \geq TU^d \times \delta$ **then**
    1- HAUUBIs.d := 1-HAUUBIs.d$\cup i_j$;
  **end if**
**end for**
$TU^U := TU^D + TU^d$;
**for** each $i_j \in$ Tq $\subseteq$ d **do**
  **if** $i_j \in$ 1-HAUUBIs.D **then**
    auub($i_j$)$^U$ := auub($i_j$)$^D$ + auub($i_j$)$^d$;
    **if** auub($i_j$)$^U \geq (TU^D + TU^d) \times \delta$ **then**
      updateADD(AUL);
      HAUUBIs.U := HAUUBIs.U$\cup i_j$;
    **else**
      updateDEL(AUL);
    **end if**
  **else**
    scan_set := scan_set$\cup i_j$;
  **end if**
**end for**
**for** $i_j \in$ scan_set **do**
  calculate auub($i_j$)$^D$;
  calculate auub($i_j$)$^U$ := auub($i_j$)$^D$ + auub($i_j$)$^d$;
  **if** auub($i_j$)$^U \geq (TU^U \times \delta$ **then**
    updateADD(AUL);
  **end if**
  **if** AUL $\neq$ **null then**
    Construct(AUL);
    update HAUUBIs.U;
  **end if**
  identify the set of HAUIs from HAUUBIs.U;
**end for**

**Figure 8.** Proposed FUP-based.

---

The AU list reduces the number of scans on the dataset and the generation of candidate itemsets. After updating the dataset, HAUUBI is added to the AU list, while non-HAUUBI is removed from the AU list. The proposed algorithm effectively updates HAUUBI to identify the actual HAUI in the updated dataset. Subsequently, the remaining itemsets in the AU list are compared against the minimum high average utility threshold, resulting in the identification of the true HAUI within the updated dataset. The proposed algorithm efficiently updates the HAUUBI to discover the actual HAUI. However, sometimes more candidate items need to be evaluated.

The FUP-HAUIMI[78] algorithm is a modified version based on the FUP concept[40], for discovering HAUI from updated datasets. The algorithm consistently preserves and updates the uncovered information, eliminating the requirement to create data for transaction deletion. Furthermore, it improves the updating process by avoiding the need for multiple scans of the dataset.

The algorithm first constructs the AU-list[39] data structure by scanning the original dataset effectively storing information for mining patterns (candidates) and gradually updating results. All items inserted in transactions are kept in the initial AU-list, and then 1-HAUUBI is classified into four categories based on the FUP concept, as described in[45], with these four categories illustrated in Fig. 9. Finally, the algorithm is able to efficiently discover updated HAUUBI and HAUI without generating candidates, as illustrated in Figs. 10, 16 and 17.

After the dataset is updated, the concept of FUP is applied to handle transaction insertions[42]. Moreover, a depth-first search approach is employed to generate candidate itemsets.

A data mining method called the FUP-HAUIMD algorithm[79], which is based on the removal of transactions from the original dataset and utilizes the MFUP (modified FUP)[40] extension from[80]. In this algorithm, deleted transactions can be categorized into four types, each with distinct implications for identifying HAUUBI (Highly Associated Unordered Unique Binary Itemsets), as illustrated in Fig. 11. In the first category, existing information can be used to determine whether the itemset remains a HAUUBI. For the second category, the item continues to be a HAUUBI. The third category can be safely discarded as it only contains non-HAUUBI. For the fourth category, a complete rescan of the original dataset is necessary. The auub value of each HAUUBI is stored in an AU list[39], and the AU list is updated every time data is removed. Mining the enumeration tree allows for the evaluation of its true HAUI without requiring multiple scans of the dataset, as illustrated in Figs. 12 and 13.

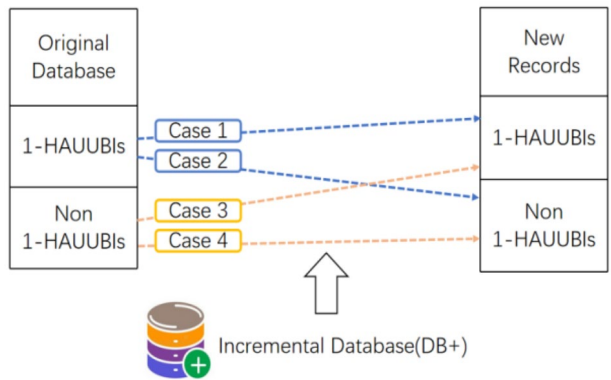

**Figure 9.** Four cases of the adapted FUP concept.

---

**Algorithm 4** FUP-HAUIMI[72] algorithm

> **Input:** $D$,the original database;
>        $utable$,a unit profit table;
>        $d$,inserted transactions;
>        $\delta$,minimum average-utility threshold;
>        $D.AULs$,the AUL-structures of $D$.
> **Output:** the set of high average-utility itemsets.
> **for** each Tq $\in$ d **do**
>      **for** each X $\in$ Tq **do**
>        build X.AUL;
>        d.AULs $\leftarrow$ ∪X.AUL;
>      **end for**
> **end for**
> **Merge**(D.AULs,d.AULs);
> **for** each X $\in$ U.AULs **do**
>      **if** $\frac{X.iu.sum}{|X|} \geq (TU^d + TU^D) \times \delta$ **then**
>        HAUIs←HAUIs∪X;
>        **if** $\frac{X.tmu.sum}{|X|} \geq (TU^d + TU^D) \times \delta$ **then**
>          exAULs←null;
>          **for** each Y afer X in U.AUL **do**
>            exAULs←exAULs+Construct(X.AUL,Y);
>            **Merge**(X,exAUL);
>          **end for**
>        **end if**
>      **end if**
> **end for**
> **for** X $\in$ U.AULs **do**
>      **if** $\frac{X.iu.sum}{|X|} \geq (TU^d + TU^D) \times \delta$ **then**
>        HAUIs $\leftarrow$ ∪X;
>      **end if**
> **end for**
>      **return HAUIs,U.AULs.**

**Figure 10.** FUP-HAUIMI algorithm.

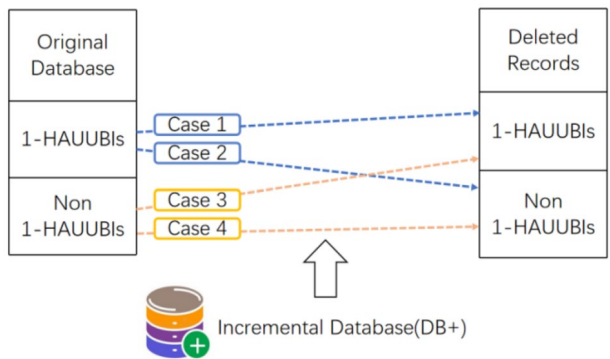

**Figure 11.** Four cases of the designed FUP-HAUIMD algorithm.

**Algorithm 5** Proposed FUP-HAUIMD[73] maintenance algorithm

**Input:** $D$,the original database;
 $d$,the deleted transactions;
 $TU^D$,the total utility in D;
 $TU^d$,the total utility in d;
 $ptable$,the profit table;
 $\delta$,the use-specified minimum high average-utility threshold;
 $D.AULs$,the AU-lists of $D$.
**Output:** $U$,the updated database(U=D-d);
 U.AULs,the updated AU-lists;
 the sets of HAUIs and HAUUBIs.
**for** each X $\in$ Tq $\wedge$ Tq $\subseteq$ d **do**
 construct X.AUL(Tq,iu,tmu);
**end for**
 d.AULs=∪X.AUL;
 **for** each X $\in$ D $\wedge$ X.AUL $\in$ D.AULs **do**
 **if** X.AUL $\neq \phi$ **then**
 search X $\in$ D.AULs $\wedge$ X $\in$ d.AULs;
 **if** $\forall$X $\in$ D.AULs $\wedge$ X $\in$ d.AULs **then**
 **for** each element $E_j$ $\in$ X.AUL **do**
 X.AUL.iu :=X.AUL.iu-$E_j$.iu;
 update X.AUL.tmu;
 X.AULs := X.AUL- $\{E_j\}$ ;
 **end for**
 U.AULs = U.AULs $\cup$ X.AUL
 **end if**
 **end if**
 **end for**
 call **DEL_Mine**($\phi$,U.AULs,$\delta$)
 **for** X $\in$ U.AULs **do**
 **if** $\sum \frac{X.AUL.iu}{|X|} \geq (TU^D - TU^d) \times \delta$ **then**
 HAUIs := HAUIs $\cup$ X;
 **end if**
 **end for**
 **return** U,U.AULs,HAUIs

**Figure 12.** Proposed FUP-HAUIMD maintenance algorithm.

Initially, Algorithm 4 scans the database to identify items from the recently added transactions, creating their AUL structures. Subsequently, the AUL structures originating from the initial database and the added transactions are combined. Upon merging the AUL structures, if the mean utility of an itemset surpasses the revised minimum average utility count, it qualifies as a high average utility itemset. Following this, its supersets are explored through a depth-first search approach based on the enumeration tree. This iterative process continues recursively until no additional tree nodes are generated. The average utility of the chosen itemsets is then computed, culminating in the algorithm's conclusion. The revised patterns are then successfully derived.

The process of Algorithm 5 commences by examining the removed transactions to form the AU-lists for 1-itemsets. Subsequently, utilizing these eliminated transactions, the AU-lists within the original database are modified, resulting in the acquisition of the revised AU-lists. Following this, Algorithm 6 is iterated recursively, merging the AU-lists of k-itemsets through a depth-first search strategy based on the enumeration tree structure. Should an itemset satisfy specific criteria, it is designated as an HAUI. In instances where these conditions are not

---

**Algorithm 6** DEL_Miner(X,$extAULOfX$,$\delta$)[73]

for each $X_a \in extAULOfX$  do
   if $\frac{X_a.AUL.iu}{|X|} \geq (TU^D - TU^d) \times \delta$ then
      HAUIs := HAUIs$\cup$ $X_a$;
      if $\sum X_a.AUL.tmu \geq (TU^D - TU^d) \times \delta$  then
         extAULOfX$_a$ := $\phi$;
         for each $X_b$ after $X_a$ in extAULOfX do
            $extAULOfX_a$:=$extAULOfX_a \cup$**Construct(X,**
            **$X_a,X_b$);**
         end for
         call **DEL_Miner(X**$_a$,**extAULOfX,**$\delta$);
      end if
   end if
end for

---

**Figure 13.** DEL_Miner algorithmin FUP-HAUIMD.

met, the auub value of the itemset is compared to the updated minimum high utility count to ascertain its superset. Additional details regarding the construction function are provided in reference[39]. Subsequent to the retrieval of the revised AU-lists, if the average utility of an itemset equals or exceeds the minimum high utility count, it is identified as an HAUI. Ultimately, the algorithm yields the updated outcomes and concludes its operation.

By default, Algorithm 7 initializes the buffer (buf) to 0 in the first iteration. Next, it computes the safety boundary (f) and the total utility d. Following this, AUL structures for all 1-item sets in d are generated to guarantee the accuracy and entirety of the resulting HAUIs. This approach is logical as, in practice, the number of transactions in d is typically small compared to the original database D. The AUL structures from D and d are then merged through a sub-routine, and the total utility of the combined databases is calculated. The updated AUL structures are managed, and if the auub value of an itemset X does not exceed the upper utility, a HAUI is detected. Subsequently, the supersets of X are evaluated for potential scanning using the recursive PRE-HAUIMI method. The list of HAUIs is updated, with PHAUIs serving as the buffer, while the AUL structures are refreshed for subsequent maintenance.

Aims to mine Highly Associated Unordered Unique Binary Itemsets (HAUI) while simultaneously reducing their search space and the number of database scans, the MHAUIPNU algorithm[81] employs a database with both positive and negative utilities. It introduces a novel, tighter upper-bound model named TUBPN, alongside a list data structure to store the required information for mining HAUI. Furthermore, three new pruning strategies are proposed to further enhance the algorithm's performance. The first strategy is based on characteristics derived from the TUBPN model, while the other two leverage attributes are associated with items (or itemsets) having negative utilities.

The paper[65] proposes an algorithm called PRE-HAUMI (High Average Utility Itemset Mining with Pre-large Itemset concept) which efficiently mines HAUI from the updated dataset with transaction insertions. The algorithm utilizes the Pre-large Itemset concept to effectively discover HAUIs and maintains an Average Utility List (AUL) structure, which ensures that each transaction is scanned at most once during the maintenance process, as illustrated in Figs. 14, 15, 16, 17.

In[63], the paper introduces an efficient algorithm called LIMHAUP, which requires only a single scan of the dataset to extract HAUP from the updated dataset, thereby reducing the cost of performing multiple dataset scans. Additionally, a new structure named HAUP List is introduced, which stores pattern information in a compact manner, eliminating the need for candidate patterns. The algorithm constructs the HAUP List through a single dataset scan and eliminates numerous irrelevant patterns, resulting in reduced execution time and memory consumption during the mining process. Initially, all HAUP Lists are rearranged in real-time order from small to large items,aiming to shrink the search space. Then, organization process is designed to rebuild the HAUP List with an effective sorting order. Ultimately, the algorithm effectively handles new insertions in the incremental dataset.

Unpromising patterns are not removed from the global HAUI list, as they might be HAUPs in a dynamic dataset. This is because the upper-bound pruning strategy can potentially overestimate the average utility. Therefore, an additional pruning strategy called MAU[82] is required to better reduce unpromising patterns. MAU rigorously mines extended patterns. The proposed algorithm demonstrates superior performance in terms of memory consumption, runtime, and scalability compared to the baseline algorithm.

The DMAUP[52] utilizes a damping window framework to extract time-sensitive patterns from incremental databases, aiming to mine high-utility frequent patterns. This method effectively extracts the latest high-utility frequent patterns, thanks to its use of damping factors to adjust item utility values based on their arrival time. Furthermore, to efficiently identify the latest high-utility frequent patterns, the method introduces new data structures known as dA-List, MU, and dUB tables. For incremental data streams, the dA-List undergoes a rebuilding process to incorporate newly added data. Moreover, the mining algorithm employs two pruning techniques, namely damping upper bound and damping maximum average utility, in compliance with the elastic properties of the damping window model. By following these steps, the method can effectively extract the most recent high-utility frequent patterns.

---

**Algorithm 7** Proposed PRE-HAUIMI[64]

**Input:** $D$,the original database;
   $utable$,an unit profit table;
   $d$,insertion transactions;
   $S^u$,upper-utility threshold;
   $S^l$,lower-utility threshold;
   $D.AULs$,the AUL-structures of D;
   $TU^D$,total utility in $D$.
**Output:** the set of high average-utility itemsets (HAUIs).
set buf ⟵ 0;
calculate the safety bound f;
calculate total utility in d as $TU^d$;
buf ⟵ buf + $TU^d$;
**for** each Tq ∈ d **do**
 **for** each X ∈ Tq **do**
  X.AUL⟵$\{Tq, iu, tmu\}$ ;
 **end for**d.AULs⟵ ∪X.AUL;
**end for**
**Merge**(D.AULs,d.AULs,U.AULs);
$TU^U$ := $TU^D$ + $TU^d$;
**for** each X ∈ U.AULs **do**
 **if** $\frac{X.iu.sum}{|X|}$ ≥ $TU^U$ × $S^u$ **then**
  HAUIs ⟵ ∪X;
  **if** X.tmu.sum ≥ $TU^U$ × $S^u$ **then**
   extAULs ⟵ null;
   **for** each each Y after X in U.AULs **do**
    extAULs ⟵ extAULs + **Construct(X.AULs,Y)**;
   **end for**
   **PRE-HAUIMI(X,extAULs)**;
  **end if**
 **end if**
**end for**
D.AULs ⟵ U.AULs;
 **return HAUIs,U.AULs,buf**

**Figure 14.** Proposed PRE-HAUIMI.

---

**Algorithm 8 Merge**(D.AULs,d.AULs,U.AULs)[64]

set X.AUL ⟵ null;
set U.AULs ⟵ null;
**for** each X ∈ d.AULs **do**
 **if** $\frac{X.tmu.sum}{|X|}$ ≥$TU^d$×$S_u$∧X∉D.AUL **then**
  **if** $TU^d$≥buf **then**
   scan D to obtain X.AUL from D;
   buf ⟵ 0;
  **end if**
  buf ⟵ buf + $TU^d$;
  **if** ∃X ∈ D.AULs ∧ X ∈ d.AULs **then**
   **for** each element $E_j$ ∈ X.AUL **do**
    X.iu.sum ⟵ X.iu.sum + $E_j$.iu;
    update X.AUL.tmu;
    X.AUL ⟵ $E_j$;
   **end for**
   U.AULs ⟵ ∪ X.AUL.
  **end if**
 **end if**
**end for**
return XY.AUL;

**Figure 15.** Merge algorithm in PRE-HAUIMI.

---

For the purpose of managing a portion of the most recent data using a sliding window model. RHUPS[83] employs an RHU list,a list-based data structure, to swiftly remove the oldest batch data from the global list, thereby displaying real-time updates of the most recent batch data in the global list. Consequently, when encountering dynamic changes in the window, the RHUPS algorithm can promptly mine the most recent efficient utility itemsets from the latest batches within the current window, without generating candidate itemsets. The data structure and mining techniques proposed in this article have the potential to develop into a large-scale machine learning system.

---

**Algorithm 9 Merge**(D.AULs,d.AULs)[72]

set X.AUL ⟵ null;
set U.AULs ⟵ null;
**for** each X ∈ d.AULs **do**
    **if** $\frac{X.iu\_sum}{|X|} \geq$ TU$^d \times \delta \wedge$ X∉D.AUL **then**
      scan D to obtain X.AUL from D;
      **if** ∃X ∈ D.AULs ∧ X ∈ d.AULs **then**
        **for** each element $E_j$ ∈ X.AUL **do**
          X.iu.sum ⟵ X.iu.sum + $E_j$.iu;
          update X.AUL.tmu;
          X.AUL ⟵ $E_j$;
          U.AULs ⟵ ∪ X.AUL.
        **end for**
      **end if**
    **end if**
**end for**
return XY.AUL;

---

**Figure 16.** Merge algorithm in FUP-HAUIMI.

---

**Algorithm 10 Construct**(X.AUL,Y)[72]

**Input:** $X.AUL$, the AUL-structures of X;
        $Y$, the itemset Y after X in X.AUL.
**Output:** $XY.AUL$, the AUL-structures of XY.
XY.AUL ⟵ null;
**if** ∃E ∈ Y.AUL ∧ X.AUL.tid == Y.AUL.tid **then**
    $E_{XY}.AUL.tid$ ⟵ $X.AUL.tid$;
    $E_{XY}.iu$ ⟵ $(X.AUL + Y.AUL)/2$;
    $update E_{XY}.tmu$;
    $E_{XY}$ ⟵ $\langle E_{XY}.tid, E_{XY}.iu, E_{XY}.tmu \rangle$;
    XY.AUL ⟵ ∪$E_{XY}$.
**end if**
return XY.AUL;

---

**Figure 17.** Construct algorithm in FUP-HAUIMI.

---

The algorithm[49] utilizes a newly developed list structure, the SHAUP list, to gather information on recent batches. By deleting the oldest batch and introducing a new one after completing the mining process of the current window, the algorithm effectively addresses the most recent stream data. The proposed approach extracts valuable and trustworthy pattern results while considering the length of pat- terns in unlimited data streams. To optimize performance, a new pruning strategy is implemented to reduce the search space, lowering the upper bound by utilizing residual utility. Prior algorithms resulted in numerous candidate patterns and suffered from performance degradation when computing the actual average utility. Conversely, our approach utilizes a list structure to store actual utility information of patterns. Through experimental analysis, results show the SHAUPM algorithm is superior in runtime, memory usage, and scalability on both real-time and synthetic datasets compared to the latest algorithms.

### Indexed list based iHAUIM

In the realm of mining high average utility patterns, multiple algorithms have been developed for handling incremental environments. Nevertheless, tree-based algorithms produce potential patterns that necessitate validation through additional database scans. Conversely, list-based algorithms do not generate potential patterns but require numerous comparison operations to identify shared transaction entries with identical identifiers throughout the mining process. These limitations have adverse impacts on algorithms aiming to expediently deliver result patterns. Conversely, indexed list structures[84,85] effectively mitigate these shortcomings and have demonstrated superior efficiency compared to tree and list structures in mining high utility patterns.

A novel method for enhancing the efficiency of current average utility driven methods is introduced in the literature as IIMHAUP[86] (Indexed List Based Incremental Mining of High Average Utility Patterns). This approach involves designing a structured list index to facilitate the mining of high average utility patterns in incremental databases. In the IIHAUP algorithm uses three key subroutines to efficiently discover resultant patterns from the initial database ODB.

## Summary and discussion
### Categories of iHAUIM
The previous section provided an overview of three primary categories of iHAUIM algorithms: those utilizing the Apriori algorithm[46,59,65,67,69], those using tree algorithms[45,55,60,75], and those relying on utility lists[44,49,52,63,65,78,79,81,83]. These algorithms differ in six key ways:

a. number of scans of the original database;
b. strategy for updating and maintaining high average utility itemsets when data changes dynamically;
c. method for searching for HAUIM;
d. type of upper bound strategy to reduce candidate itemsets;
e. type of data structure for maintaining transaction and itemset information (tree-based or utility-list-based);
f. pruning strategies to reduce search space and speed up mining.

Tables 6, 7 summarizes these characteristics for the 19 algorithms discussed, noting that not all have been comprehensively studied in the literature. Moving forward, we will delve deeper into these iHUIM algorithms, analyzing and discussing them from the angles of runtime and memory consumption.

### Runtime, memory consumption and scalability
The performance of various algorithms for itemset mining has been evaluated, including those proposed by APITPAU Hong et al.[67] and SHAU Yunet al.[61] that utilize tree structures, as well as IHAUPM Lin et al.[45], FUPHAUIMI Zhang et al.[78], and LIMHAUP Kim et al.[63] that use utility lists. The results indicate that utility-list-based algorithms exhibit superior performance comparable to Apriori-based methods. Each iHUIM algorithm has its own limitations, which have been analyzed. Both utility-list-based and tree-structure-based approaches can reduce the number of candidate itemsets generated and the transactions scanned during maintenance. The use of the pre-large concept strategy has been found to be more effective than the FUP concept strategy based on experimental results obtained from FUP-based Wu et al.[44] and PRE-HAUIMI Lin et al.[65]. Lastly, sliding windows and pruning techniques have been shown to enhance the runtime of the algorithm based on the experimental results of LIMHAUP Kim et al.[63] and SHAUPM Lee et al.[49].

| Type | Algorithm | Test datasets | Compared algorithms | Notes |
|---|---|---|---|---|
| Apriori-based iHAUIM | ITPAU[67], (2009) | A real data was from a major grocery chain store in America | TPAU[68] | ITPAU is a two-stage algorithm based on the FUP concept, employing hierarchical search for HAUI and the Apriori method |
| | M-TP[59], (2011) | T10I4N4KD200K | Batch-TP[68] | Similar to the Apriori algorithm, this approach involves multiple scans of the dataset and generates numerous unpromising candi- date items |
| | PRE-HAUI-DEL[69], (2021) | Mushroom Foodmart BMS Accidents Chess Retail | Apriori[23] Apriori(Ipub) | The pre-large concept is applied to HAUIM and is used to remove transactions in dynamic databases Additionally, the Ipub upper bound model is applied, which can significantly reduce the number of checked candidate items in the search space |
| | APHAUI[65], (2020) | Retail Foodmart BMS Mushroom Chess Accidents | Apriori(A) Apriori(A,Ipub) | The authors proposed an Apriori-based pre-large algorithm,APHAUI, which uses a linked list structure to maintain transactions and requires at most one scan of each transaction during the entire maintenance process |
| | APHAUIM[46], (2020) | Retail Mushroom | ITPAU[67] | Apriori and the pre-large concept |
| Tree-based iHAUIM | SHAU[61], (2016) | Chain-store Retail Mushroom Chess | STPAU[37] ITPAU[67] | This method is utilized to retain information from recent streaming data and employs a strategy known as RUG to decrease the number of generated candidate items |
| | EHAUI-Tree[75], (2017) | Accident Retail | HAUI-Tree[76] | It employs a data structure to preserve itemsets, enabling it to mine HAUI from the updated dataset without the need for restarting |
| | IHAUPM[45], (2018) | Foodmart Kosarak Mushroom Retail T10I4D100k T40I10D100K | PAI[38] TPAU[36] HAUI-tree[76] HAUI-Miner[39] HAUP-growth[8] | The algorithm utilizes the FUP concept to incorporate transactions from incremental datasets Uses an efficient HAUP tree structure |
| | IMHAUI[60], (2017) | Chain-store Foodmart Mushroom Breast-cancer Wisconsin | ITPAU[67] | The path adjustment method was modified to reconstruct the IHAUI-tree based on the descending order of AUUB |
| | MAM[55], (2018) | Breast cancer Wisconsin Liver disorders Heart cleveland Hepatitis | ITPAU[67] UP-Growth*[33] IMHAUI[60] | The algorithm utilizes DAT-tree and TUL-list data structures |

**Table 6.** IHAUIM algorithm.

| Type | Algorithm | Test datasets | Compared algorithms | Notes |
|------|-----------|---------------|---------------------|-------|
| List-based iHAUIM | FUP-based[44], (2017) | T10I4D100K<br>Accidents | HAUI-Miner[39] | The algorithm employs an AU-list, which help maintain and reduce the cost of multiple database scans without generating a large number of candidates |
| | FUP-HAUIMI[78], (2018) | Retail<br>T10I4D100K<br>Kosarak<br>T10I4N4KD100K<br>Mushroom<br>Foodmart | HAUI-Miner[39]<br>IHAUPM[45] | The FUP-HAUIMI algorithm has been updated by utilizing the FUP concept to effectively save the information of the mining patterns using the AUL structure |
| | FUP-HAUIMD[79], (2018) | T10I4N4KD100K<br>Accidents<br>Foodmart<br>Mushroom<br>Retail | HAUI-Miner[39]<br>EHAUPM[87] | Utilizes AU-list and the modified MFUP concept to maintain the discovered HAUIs |
| | MHAUIPNU[81], (2019) | Chess<br>Mushroom<br>Accidents<br>Pumsb<br>Retail<br>Kosarak | Na¨Ivetubpn[81]<br>Na¨auubpnIve[81] | The algorithm employs TUBPN to reduce the search space for mining HAUIs |
| | PRE-HAUIMI[65], (2020) | T10I4D100K<br>Retail<br>Kosarak<br>T40I10D100K<br>Mushroom<br>Foodmart | FUP-based[44]<br>IHAUPM[45]<br>HAUI-Miner[39] | The algorithm utilizes the AUL-list structure |
| | LIMHAUP[63], (2020) | Retail<br>Chess<br>Foodmart<br>Mushroom<br>T10I4DxK<br>Tx1Nx2Lx3 | ITPAU[67]<br>IMHAUI[60] | The HAUP-list was adopted, which compactly stores information about patterns and easily removes many hopeless patterns |
| | DMAUP[52], (2021) | Chain-store<br>Kosarak<br>Accidents<br>Pumsb | MPM[55]<br>GENHUI[88]<br>I-MHAI[82] | For incremental data streams, the dA-List undergoes a rebuilding process to incorporate newly added data |
| | RHUPS[83] (2021) | T10I4DxK<br>Tx1Nx2Lx3 | SHUPM[89]<br>DSHUP | The RHUPS algorithm applies a list-based data structure, sliding window, and time decay concept |
| | SHAUPM[49] (2022) | Chain-store<br>Chess<br>Retail<br>Foodmart<br>T10I4DxK<br>Tx1Nx2Lx3 | SHAU[61]<br>LIMHAUP[63] LMHAUP[90]<br>HAUI-Miner[39] | The algorithm extracts the entire set of recently discovered high average utility patterns without creating candidates in a single pass through the streaming data |

**Table 7.** IHAUIM Algorithm.

## Challenges and future directions

Despite the effectiveness of the existing methods, there are still many future directions that require being explored. Following are some crucial research opportunities associated with the iHAUIM algorithm.

*Enhancing the effectiveness of the algorithms*
The iHAUIM algorithm can be time-consuming and occupy a large memory while executing, which can raise concerns in real-time dynamic database updates. Even though the current incremental high-utility mining algorithms are faster than their predecessors, there is a scope for improvement. To name a few, compact data structures like trees or lists and more efficient pruning strategies could be developed for mining methods.

*Handling the complex dynamic data*
Real-life data is highly dynamic, comprising vast and complex datasets used in various fields. Although the principle behind it is straightforward, integrating it into the design of data mining algorithms is complicated. Discovering dynamic data environments is much more difficult and challenging than analyzing static data.

*Analyzing the massive amounts of data*
Incremental mining of big databases has higher computational costs and memory consumption. Nonetheless, in the era of big data, processing data step-by-step and having a look at earlier analyzed results is indispensable. Research opportunities exist for iHAUIM to process large databases, such as designing parallelized iHUIM algorithms.

*Analyzing the runtime*

In the experiment, we assessed the runtime of five algorithms across various TH values while maintaining a fixed IR (=1%), as depicted in Fig. 18. As depicted in Fig. 18, it's clear that the designed PRE-HAUIMI algorithm outperforms the other two algorithms across six datasets.

As the TH value increases, the running time of the five algorithms decreases. This is reasonable because as TH increases, less HAUI is found. Therefore, these five algorithms require less runtime. In addition, it can be seen that for some datasets, such as Fig. 19a,c,f, the PRE-HAUIMI algorithm designed remains stable for various TH values. HAUI Miner represents the most advanced algorithm for mining HAUI using the auub model, while IHAUIM stands as the most advanced algorithm for incremental HAUIM utilizing tree structures. Consequently, it can be concluded that the designed PRE-HAUUIMI, FUP-HAUIMI, and FUP-based algorithms exhibit strong performance when handling dynamic databases with transaction inserts. The efficiency of the AUL (Average

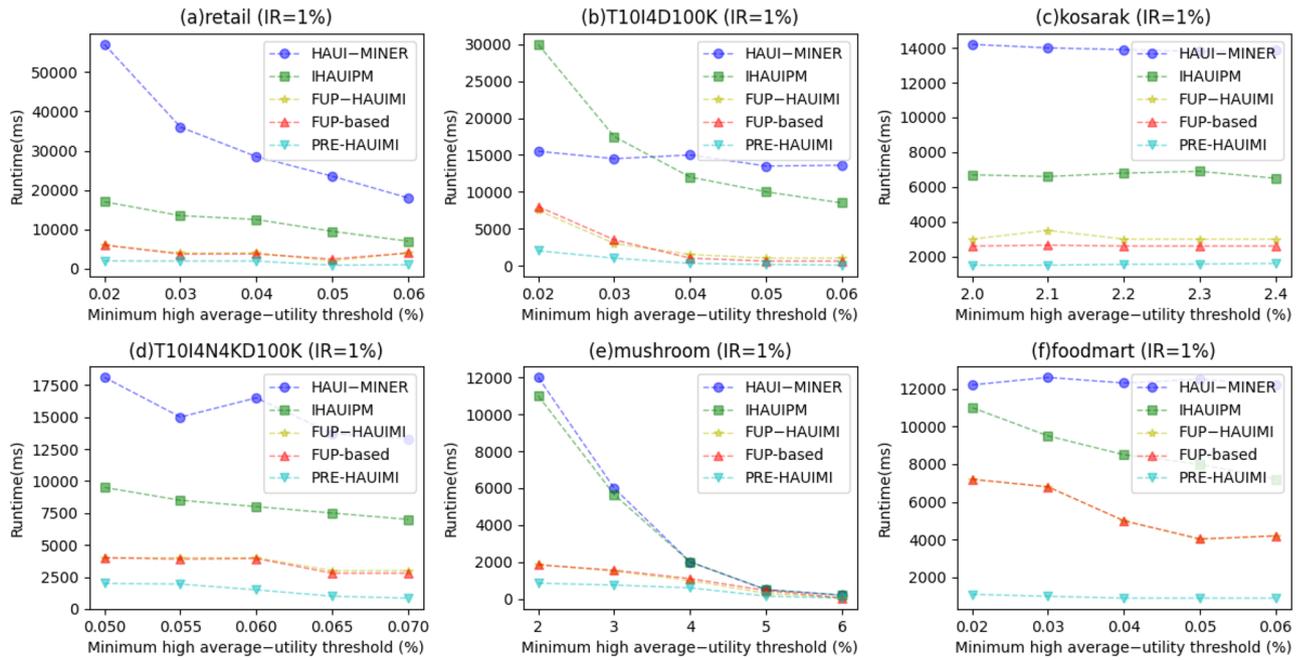

**Figure 18.** Runtimes for various threshold values.

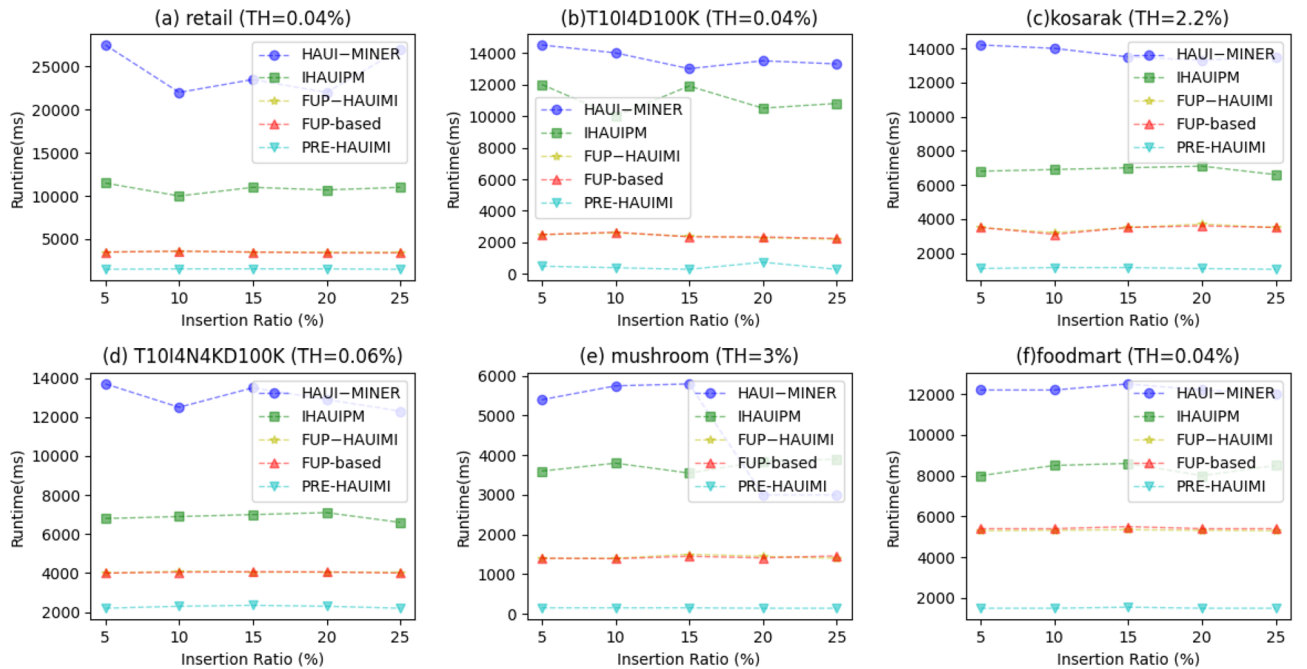

**Figure 19.** Runtimes for various insertion ratios.

Utility List) structure facilitates streamlined calculations and retrieval of the required HAUI. Experimental evaluations were conducted on six datasets, maintaining fixed TH (Transaction-Utility) values, while varying IR (Item Reduction) values. Figure 19 presents the results derived from these experiments, showcasing the comparative performance of the algorithms.

As illustrated in Fig. 19, the PRE-HAUIMI algorithm demonstrates superior performance compared to both FUP-HAUIMI and FUP-based algorithms. Furthermore, it is observed that the FUP-HAUIMI and FUP-based algorithms still outperform the HAUI Miner and IHAUPM algorithms. The stability of all algorithms, particularly the PRE-HAUIMI algorithm, is evident as the IR (Item Reduction) increases. This indicates that as the IR increases, the performance of all algorithms remains consistent, with the PRE-HAUIMI algorithm consistently displaying the best performance.

*Memory usage improvement*
We conducted experiments to analyze the memory usage of various algorithms considering fixed IR values and different TH values. The results are depicted in Fig. 20. Notably, the HAUI Miner algorithm demonstrates superior memory usage performance across datasets (Fig. 20a,c,e). This can be attributed to the utilization of a utility list structure in HAUI Miner, which efficiently compresses and maintains discovered information. As a result, it usually demands less memory when compared to the IHAUPM algorithm, which utilizes a tree structure for incremental maintenance. Moreover, HAUI Miner doesn't necessitate holding extra information for maintenance purposes. Instead, when the database size changes, the algorithm rescan the database to acquire updated information, resulting in potential computational costs but lesser memory requirements.

Through experiments with fixed IR values and different TH values, we evaluated the memory usage of various algorithms. Figure 21 illustrates the results, showcasing the superior memory usage performance of the HAUI Miner algorithm across datasets 21a, c, and e. This advantage can be attributed to the efficient compression and maintenance of discovered information facilitated by the utility list structure utilized by HAUI Miner. Consequently, it requires less memory compared to the IHAUPM algorithm, which employs a tree structure for incremental maintenance. Additionally, HAUI Miner does not require the retention of additional information for maintenance. Instead, it rescans the database when its size changes, obtaining updated information at the cost of computational overhead but with reduced memory requirements.

*Number of patterns*
The experiment involved evaluating the number of candidate patterns generated during the discovery of actual HAUI. The results, considering different TH values with fixed IR, are presented in Fig. 22. Observing Fig. 22, it is evident that, with the exception of Fig. 22c and d, the proposed PRE-HAUIMI, FUP-HAUIMI, and FUP-based algorithms generate significantly fewer candidate patterns compared to the HAUI Miner and IHAUPM algorithms. Notably, the PRE-HAUIMI algorithm produces the fewest number of candidate patterns.

This discrepancy can be attributed to the dense nature of the T10I4N4KD100K dataset, where many transactions contain the same maintenance items. As a result, the proposed PRE-HAUIMI, FUP-HAUIMI, and FUP-based algorithms may require additional checks in the enumeration tree to determine if a superset needs to be generated. However, overall, these algorithms still evaluate fewer patterns compared to the other algorithms. This highlights the effectiveness of the AUL structure and adaptive FUP (Frequent Utility Pattern) concept in reducing the incremental mining cost of average utility itemsets. The results, considering different DR (Dependency Ratio) values with fixed TH, are depicted in Fig. 23.

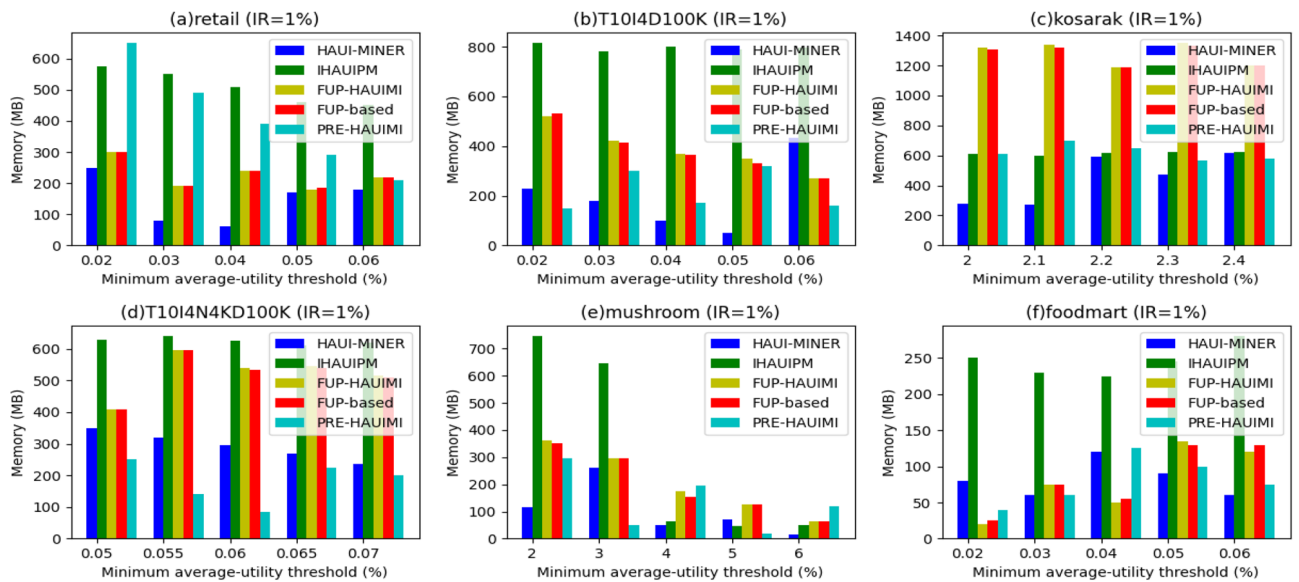

**Figure 20.** The results of memory usage w.r.t varied thresholds.

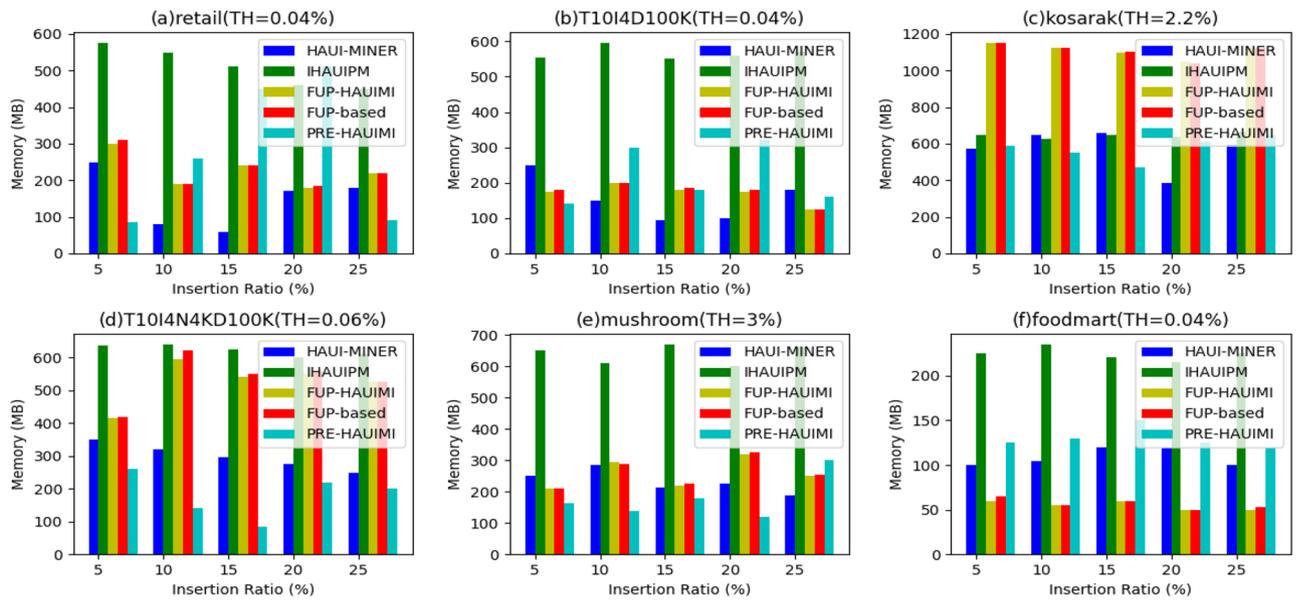

**Figure 21.** Usage for various insertion ratios.

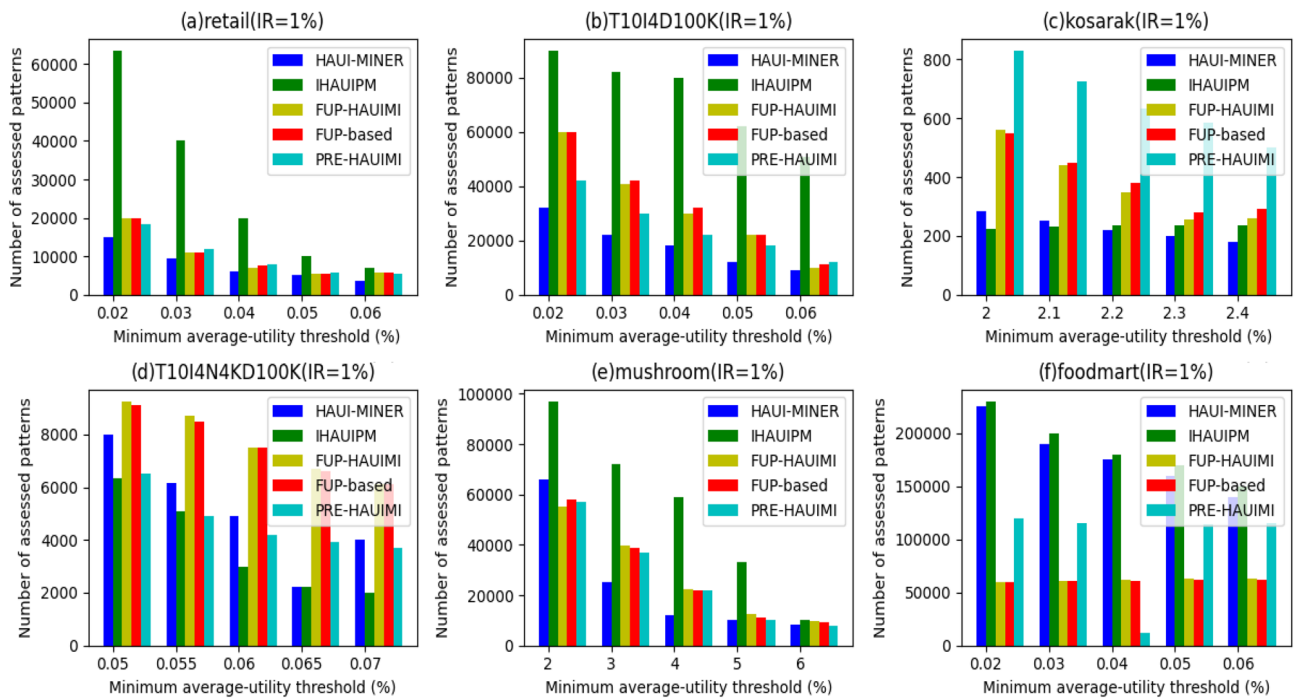

**Figure 22.** Number of candidate patterns for various threshold values.

Similarly, it is observed that in very sparse and dense datasets, such as the ones depicted in Fig. 23c and d, the PRE-HAUUIMI, FUP-HAUIMI, and FUP-based algorithms may require checking more candidate patterns. However, for other datasets, like those in Fig. 23a,b,e, these algorithms surpass the performance of the IHAUPM algorithm and even achieve the best outcomes, as demonstrated in Fig. 23f.

In terms of runtime performance, the proposed PRE-HAUUIMI, FUP-HAUIMI, and FUP-based algorithms outshine the alternative approaches. This can be attributed to the efficiency derived from the FUP concept and the AUL structure, enabling a significant reduction in runtime. Considering the overall results, it can be inferred that while the PRE-HAUUIMI, FUP-HAUIMI, and FUP-based algorithms require additional memory usage and may need to check more candidate patterns in certain scenarios, nevertheless, they consistently achieve higher levels of efficiency and effectiveness in the majority of cases. Among them, the PRE-HAUUIMI algorithm performs the best, with the exception of very sparse datasets with long transactions or extremely dense datasets.

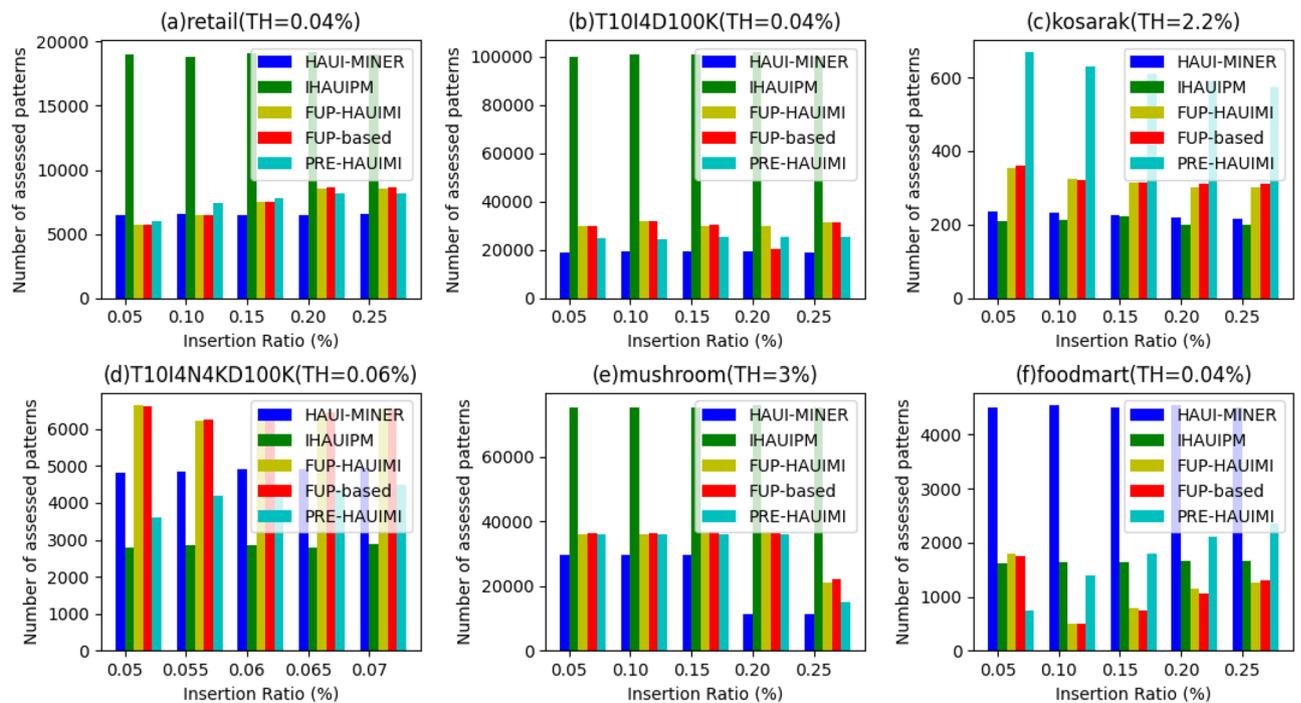

**Figure 23.** Number of candidate patterns for various insertion ratios.

Taking these findings into account, it becomes evident that there are numerous directions that can be explored to further enhance and improve the iHAUIM algorithm, catering to the ever-evolving and dynamic demands of data mining.

## Conclusion

A detailed summary of different algorithms for the IHAUIM problem is presented in this paper. We provide an all-inclusive and current analysis of IHAUIM algorithms in dynamic datasets and propose a classification system for the existing IHAUIM techniques. We explore various iHAUIM algorithms for modifying datasets in dynamic data settings, streaming data, and sequential datasets, and evaluate the advantages and drawbacks of the most advanced approaches. Additionally, we identify the significant areas for future research in incremental high-average utility itemset mining.

### Data availability

The dataset used in this paper is a publicly available dataset sourced from the internet, and it can be accessed from the following website: https://www.kaggle.com/uciml/pima-indians-diabetes-database.

## Acknowledgements

The subject is sponsored by the National Natural Science Foundation of P. R. China (No.61976120, No. 62102194, No. 62102196). Natural Science Foundation of Inner Mongolia Autonomous Region of China (No. 2022MS06010), Natural Science Research Project of Department of Education of Guizhou Province (No. QJJ2022015). Inner Mongolia Autonomous Region Higher Education Institutions Science and Technology Research Project (NJSY23004). Scientific Research Project of Baotou Teachers' College (BSYKY2021-ZZ01, BSYHY202212, BSYHY202211, BSJG23Z07).

## Author contributions

Chen Jing: writing original draft, review response, commentary, revision. Yang Shengyi: writing original draft, review response, commentary, revision. Weiping Ding and Li Peng: conceptualization, funding acquisition, methodology, supervision, review. Liu Aijun: writing original draft, commentary. Zhang Hongjun: validations. Tian Li: validation, conceptualization.

## Competing interests

The authors declare no competing interests.

## Additional information

**Correspondence** and requests for materials should be addressed to W.D. or A.L.

**Reprints and permissions information** is available at www.nature.com/reprints.

**Publisher's note** Springer Nature remains neutral with regard to jurisdictional claims in published maps and institutional affiliations.